# Inferring the main drivers of SARS-CoV-2 global transmissibility by feature selection methods

**Marko Djordjevic[1,*], Igor Salom[2], Sofija Markovic[1], Andjela Rodic[1], Ognjen Milicevic[3], Magdalena Djordjevic[2]**

[1]Quantitative Biology Group, Institute of Physiology and Biochemistry, Faculty of Biology, University of Belgrade, Serbia

[2]Institute of Physics Belgrade, National Institute of the Republic of Serbia, University of Belgrade, Serbia

[3]Department for Medical Statistics and Informatics, School of Medicine, University of Belgrade, Serbia

**\* Correspondence:**

Marko Djordjevic, e-mail: dmarko@bio.bg.ac.rs

## Abstract

Identifying the main environmental drivers of SARS-CoV-2 transmissibility in the population is crucial for understanding current and potential future outbursts of COVID-19 and other infectious diseases. To address this problem, we concentrate on basic reproduction number $R_0$, which is not sensitive to testing coverage and represents transmissibility in an absence of social distancing and in a completely susceptible population. While many variables may potentially influence $R_0$, a high correlation between these variables may obscure the result interpretation. Consequently, we combine Principal Component Analysis with feature selection methods from several regression-based approaches to identify the main demographic and meteorological drivers behind $R_0$. We robustly obtain that country's wealth/development (GDP per capita or Human Development Index) is the most important $R_0$ predictor at the global level, probably being a good proxy for the overall contact frequency in a population. This main effect is modulated by built-up area per capita (crowdedness in indoor space), onset of infection (likely related to increased awareness of infection risks), net migration, unhealthy living lifestyle/conditions including pollution, seasonality, and possibly BCG vaccination prevalence. Also, we argue that several variables that significantly correlate with transmissibility do not directly influence $R_0$ or affect it differently than suggested by naïve analysis.

## 1    Introduction

Despite the unprecedented worldwide campaign of mass immunization, due to the relatively slow vaccine rollout and to the appearance of new, more contagious (Tegally et al., 2020), and maybe even more deadly SARS-CoV-2 strains (Mallapaty, 2021), COVID-19 still takes its toll on human lives, stifles the world economy, and forces the majority of countries to keep unpopular lockdowns. In the absence of a prompt solution to the first pandemic of the century, the goal to identify the main environmental and demographic parameters that influence the dynamics of infection transmission remains as important as ever.

Investigating which factors influence the rate of COVID-19 expansion is a highly involved matter, primarily due to many possibly relevant modes of SARS-CoV-2 transmission. It is generally undisputed that COVID-19 transmission occurs by droplets between people in close proximity (Rahimi et al., 2021). Consequently, various factors that influence the frequency of interpersonal contacts and intensity of social mixing between people arriving from different/distant areas will necessarily strongly impact the rate of the epidemic spread. However, other means of transmission may also play important roles. It has been established that contact with surfaces in the immediate environment of infected persons (or with objects used by them) can lead to infection. Also, there is an increasing body of evidence that the airborne transmission of COVID-19 cannot be neglected and, moreover, that small air pollutant particles may aid the long-range virus transmission (Coccia, 2020b; Rahimi et al., 2021). Some authors even suggest the relevance of contaminated wastewater in the infection spread (Rahimi et al., 2021). Such various modes of transmission open up the possibility not only of direct influence of numerous environmental factors (e.g., temperature, humidity, wind speed, air pollution) (Sarkodie and Owusu, 2020) but also of nontrivial interactions between these factors (for example, the interplay between the pollution, wind speed and population density is discussed in (Coccia, 2021b)). It must be also taken into account that the level of personal susceptibility to the given virus can strongly affect the rate of an epidemic, and thus a comprehensive study must also consider more subtle demographic/medical variables (Notari and Torrieri, 2020).

We recently published a comprehensive study of the correlation of 42 different demographic and weather parameters with COVID-19 basic reproduction number $R_0$ across 118 world countries (Salom et al., 2021). $R_0$ is a well-established epidemiological measure of virus transmissibility. Its major advantage is independence on the testing policy/capacity and intervention measures, which can be highly variable (and almost impossible to consistently control) between different countries (Salom et al., 2021). In (Salom et al., 2021), we selected all the countries that exhibited regular exponential growth in the case numbers before the introduction of intervention measures (Djordjevic et al., 2021), from which their $R_0$ values can be reliably extracted. Tracking a wide range of countries allows achieving a maximal variability in the dataset, i.e., a maximal possible range in the values of analyzed variables, as another advantage of this study. This dataset will be used as a starting point in this work.

Our present goal is to go a step further than this previous study by applying and combining various advanced data analysis methods. Namely, while (Salom et al., 2021) covered a broad scope of variables and countries, it focused on establishing pairwise correlations between $R_0$ and each of the studied factors, ignoring the fact that many of these variables are highly mutually correlated. This is most obvious in the case of the weather parameters such as e.g. temperature and UV radiation (which both reflect the local climate in a similar way and follow comparable seasonal trends), but also in the case of many demographic parameters, e.g. the strong positive correlation between the Human Development Index (HDI) and cholesterol levels. Based on pairwise correlations alone, it is thus hard to estimate which of these variables might be truly influencing the spread of the disease, to what extent, and in which direction.

Therefore, in the present paper, our task is to go beyond mere pairwise correlations and, by using a combination of principal component analysis (PCA) and four different regression-based approaches, establish which of these factors significantly affect $R_0$ (by at the same time controlling for the impact of other factors). To achieve this, the number of variables necessary to explain the virus transmissibility needs to be reduced to only a few without losing predictiveness. However, this is not the only challenge, because of variable redundancy. In particular, one may select different combinations of variables accounting together for a similar proportion of variance in the virus transmissibility, which presents an ambiguity that is difficult to resolve. Consequently, there is a challenge to narrow down the possibilities and illuminate important contributions of the seemingly small differences between highly correlated variables. Notably, numerous studies examined the correlations of several selected (Lin et al., 2020; Ran et al., 2020; Xie et al., 2020) or many different (Li et al., 2020; Hassan et al., 2021; Salom et al., 2021) sociodemographic and meteorological factors with the magnitude of the COVID-19 epidemic. However, only a few studies tried to select a handful of key factors whose combination can explain a large portion of the variance between regions (Allel et al., 2020; Coccia, 2020d; Gupta and Gharehgozli, 2020; Notari and Torrieri, 2020). Even a smaller number of studies included data from multiple countries (Allel et al., 2020; Notari and Torrieri, 2020). Furthermore, a similar effort to relate the rate of exponential case growth to a smaller set of variables by PCA was made (Notari and Torrieri, 2020; Notari, 2021). To conduct a thorough investigation and reach robust conclusions, we partially decorrelate 24 different factors and apply several specialized feature selection methods independently. While disentangling the various effects on the epidemic spread is undeniably a challenging task, by comparing the results obtained by different methods one can identify the variables that are selected by all or most of them, which provides valuable evidence in favor of their true significance.

The main idea of this study is to develop a novel approach, which can robustly identify the most important predictors of $R_0$. The development of such an approach will *i)* provide a straightforward solution to the known problem of selecting important among the highly correlated variables, *ii)* enable a better understanding of which environmental and demographic variables may dominantly and/or independently influence the progression of the COVID-19 epidemics, and what is the direction of this influence. With these goals, we organize the study as follows:

1. The variables are first naturally split into two groups. The first group comprises six meteorological parameters, sampled and averaged (for each country) during the initial stage of the local epidemic outbreak: air temperature (T), precipitation (PR), specific humidity (H), ultra-violet radiation index (UV), air pressure (P), and wind speed (WS). Eighteen (broadly-speaking) demographic parameters form the second group: human development index (HDI), percentage of the urban population (UP), gross domestic product per capita (GDP), amount of the built-up area per person (BUAPC), percentage of refugees (RE), net migration (i.e., the number of immigrants minus emigrants, I-E), infant mortality (IM), median age (MA), long-term average of PM2.5 pollution (PL), prevalence and severity of COVID-19 relevant chronic diseases in the population (CD), average blood cholesterol level (CH), the prevalence of raised blood pressure (RBP), the prevalence of obesity (OB), the prevalence of insufficient physical activity among adults (IN), BCG immunization coverage (BCG), alcohol consumption per capita (ALC), smoking prevalence (SM), and the delay of the epidemic onset (ON).

2. Due to strong correlations between parameters within each group (as well as across the groups, but at a lower extent), on each group, we will perform the PCA (Jolliffe, 2002). This step will allow us to notably reduce the dimensionality, i.e., proceed to work with a smaller number of (mostly) uncorrelated variables. Such dimensionality reduction will significantly simplify the further analysis and improve the reliability of the results.

3. The linear regression analysis will next be performed in four independent ways, ranging from our custom-developed to more formal regression-based approaches, to select important variables. In our custom-developed approach, multiple linear regressions are applied, first separately to demographic and meteorological principal components (PCs) to narrow down the number of relevant PCs within each of the two groups, before doing overall linear regression with the remaining PCs to assess their importance in explaining $R_0$. A major advantage of such analysis is an intuitive understanding of the data structure and its relation to $R_0$. This analysis is next independently redone by more formal feature selection methods, commonly employed in bioinformatics and systems biology: Stepwise regression and regressions utilizing both regularization and variable selection - Lasso (Least Absolute Selection and Shrinkage Operator) and Elastic net (Tibshirani, 1996; Zou and Hastie, 2005; Hastie et al., 2009). Lasso and Elastic net are regressions based on regularization, which can shrink coefficients exactly to zero, allowing variable selection. Such feature selection is quite important, as the variables that do not affect the response ($R_0$) may introduce significant noise in the model. This would lead to overfitting (high variance), which is exacerbated by a small/limited dataset and correlated predictors (as applicable here). Note that regressions with regularization are not ordinary fits, e.g. hyperparameters that control coefficient shrinkage have to be carefully tuned through cross-validation. Overall, this comprehensive analysis will ensure the consistency and robustness of the reported results.

4. Finally, an intuitive interpretation of the obtained results will be presented. This will permit a much more specific understanding of COVID-19 transmissibility, by focusing on the main driving factors behind the disease spread in the population.

## 2    Methods

### 2.1    Sample and data

Data for demographic and meteorological parameters were assembled as described in (Salom et al., 2021). Briefly, the data correspond to six meteorological and eighteen demographic variables outlined above. The differences between this dataset and the one used in (Salom et al., 2021) is the following: IMS (Social security and health insurance coverage), Prevalence of ABO and Rhesus blood groups, and Ambient levels of different pollutants ($NO_2$, $SO_2$, CO, PM2.5, PM10) are not used in this analysis, as they contain too many missing values. Instead of the pollutant levels measured from air pollution monitoring stations during the epidemic's exponential growth (available for only ~40 countries), we use the yearly average PM2.5 pollutant levels in 2017 (World Bank, 2020b). Also, we consider GDP per capita (GDPpc), taken from (World Bank, 2020a), as a more direct (average) indicator of a country's economic wealth/productivity.

There are no missing values in the meteorological data, while we substitute the missing values in the demographic data (which were sparse for the used variables) with the median values of the respective variables. The discarded variables have at least 30% missing values (occurring for blood groups) and going up to 72% (for CO). On the other hand, for the variables that we retained, the missing values are sparse. Specifically, the maximal fraction is 6% (for refugees).

Basic reproduction number ($R_0$), i.e., a measure of SARS-CoV-2 transmissibility in a fully susceptible population and in the absence of intervention measures (social distancing, quarantine), was also taken from (Salom et al., 2021), where it was inferred from non-linear dynamics modeling. Overall, demographic data, meteorological data and $R_0$ were assembled for 118 different countries from which we could reliably infer $R_0$. We used the following criteria to select the countries for which $R_0$ was inferred and consequently used in the further analysis:

i)    We initially selected 165 world countries, with the following criteria: more than 1000 tests performed by 12.06.2020, with the ratio between performed tests and detected cases larger than

5. This accounts for reliable testing capacity, i.e., in this way, we avoid countries with low testing capacity.

ii) $R_0$ is inferred from the regime which has reached the deterministic limit, i.e., the starting time is such that the number of detected cases is ~10 or more. Each country is processed/inspected manually to determine the time period that is used in the analysis (i.e., exactly which interval corresponds to the exponential growth) - this is necessary as we found that this interval significantly varies for different countries. The manual inspection corresponds to recognizing a straight line on a semi-logarithmic plot, which is straightforward and reproducible.

iii) Clear exponential growth is observed for at least seven days, corresponding to the straight line in the number of detected cases vs. time on the log scale.

In this way, we avoid the cases where too small (or irregular) testing is preventing reliable inference of $R_0$. Two `filters' insure this: The first is a condition *i*) which directly accounts for those countries with too small testing capacity. Secondly, the countries with `irregular' testing are also filtered by condition *iii*), as they exhibit an irregular behavior in the case count numbers, rather than regular exponential growth, which is robustly observed across other countries. Finally, condition *ii*) ensures that stochastic fluctuations (and possible very early testing inconsistencies) do not influence results, while the manual inspection for each country ensures adequacy of the time interval used in the analysis.

## 2.2 Measures of variables

Several variables, particularly among demographic data, show a significant deviation from normality when visually inspected. Such deviations generate large outliers and would significantly impact the necessary normality of the model error residuals. We consequently transform the data (where necessary) to make the resulting distributions closer to normal, by using standard transformations that reduce the right and left skewness. We chose the strength of the applied transformations (e.g., square root, cubic root, or log) so that skewness of the transformed distribution is as close as possible to zero. Applied transformations are provided in Table 1. Each transformed variable whose direction was changed by the transformation was taken with a minus sign, so that the original and the transformed variable are oriented in the same direction, allowing for easier result interpretation.

After transformations, the remaining (now sparse, less than 2% of the dataset) outliers were removed by substituting them with the median of each variable; the outliers were identified as having more than three scaled median absolute deviation (MAD) from the (transformed) variable median. Removal of the outliers is important, as they may substantially (both quantitatively and qualitatively) obscure the multivariate regression analysis, including regressions with regularization (Hastie et al., 2009).

The dimensionality of the transformed data was next reduced and the data decorrelated through PCA (Jolliffe, 2002). PCA was done separately for demographic and meteorological variables to allow for a more straightforward interpretation of the obtained PCs. Since different variables are expressed in different units and correspond to diverse scales, each variable in the dataset was standardized (the mean subtracted and divided by the standard deviation) before PCA. For both datasets, we retained as many PCs (starting from the most dominant one) as needed to (cumulatively) explain >85% of the data variance. It was inspected that PCs reasonably follow a normal distribution (as expected, based on the transformation of the original variables). Few remaining outliers were then substituted by medians. For easier interpretation of PCs and their contribution to $R_0$, each PC was oriented in the same direction as the variable with which it has a maximal magnitude of Pearson correlation (i.e., when needed, the sign of the PC was flipped to render the positive sign of this correlation).

### *2.3   Data analysis*

*Custom regression analysis* was done by applying multiple linear regression (PC regression) to only demographic PCs (Hastie et al., 2009). Only linear terms were included in the regression to allow straightforward interpretation, i.e., selection of PCs that significantly affect $R_0$. Significant PCs were selected as those appearing in the regression with P<0.05, where the significance in the regression was estimated in the standard way (through F-statistics) (Alexopoulos, 2010). The same regression was then repeated with only meteorological PCs, and those significant in explaining $R_0$ were retained. Finally, multiple linear regression was performed with all retained demographic and meteorological PCs. The significant PCs from this last step were recognized as PCs relevant for the $R_0$ explanation. Before regression, each PC was standardized so that coefficients obtained in the regression provided a measure of the variable importance in explaining $R_0$. For both the custom analysis and stepwise regression, OLS (Ordinary Least Squares) were used as the regression metrics.

*Stepwise regression* was used to select PCs that significantly affect $R_0$. In Stepwise regression, as well as in LASSO and Elastic net described below, all PCs (demographic and meteorological) were included in the regression. Briefly, starting from a constant model, at each step a term is added to the model if its significance (calculated with F-statistics) meets the condition P<0.05 (Pope and Webster, 1972). Only linear terms are added to the model (i.e., interaction and quadratic terms are not considered) to allow for straightforward interpretation which PCs significantly affect $R_0$. All PCs are standardized before regression so that contributions of the terms (PCs) in the model can be assessed by the magnitude of the regression coefficient.

*Lasso regression* (Tibshirani, 1996; Hastie et al., 2009) was used to implement L1 regularization. As needed with the Lasso regularization, all PCs were standardized before regression, which allowed direct comparison of the coefficients obtained by the regression. The value $\lambda$ in Lasso was treated as the hyperparameter, i.e., $\lambda_{min}$ value was determined through cross-validation, so that MSE (Mean Squared Error) on the testing set was minimal. A total of 100 $\lambda$ values were put on the grid, corresponding to the geometric sequence, where the largest value produces all zero terms. Note that larger $\lambda$ corresponds to a sparser model, i.e., a smaller number of non-zero components in the regression, while the small $\lambda$ limit corresponds to OLS regression. To obtain the maximally sparse model, $\lambda_{1SE} = \lambda_{min} + 1SE$, where $1SE$ corresponds to the standard error of MSE obtained by cross-validation, was used. 1000 cross-validations were performed - in each repetition, 20% of the data were randomly selected for the testing set, with the remainder used for training. All non-zero terms, and the corresponding coefficients obtained through Lasso, were reported.

*Elastic net regression* was used to implement a combination of L1 and L2 regularization (Zou and Hastie, 2005). Analogously to our Lasso analysis, i.e., as needed due to regularization, all PCs were standardized. In the regression, both $\alpha$ and $\lambda$ were treated as hyperparameters, i.e., their optimal values were found by cross-validation. Cross-validation was repeated 1000 times. In each repetition, testing and training sets were formed in the same way as for Lasso. $\alpha$ and $\lambda$ values were put on a grid consisting of 100 $\alpha$ and 100 $\lambda$ values. $\alpha$ values on the grid were chosen uniformly in the range [0,1] - $\alpha$ approaching zero corresponds to Ridge (L2) regression, and 1 corresponds to Lasso regression. For each $\alpha$ value, $\lambda$ values were chosen as described for the Lasso regression. For each repetition of cross-validation, a combination of $\alpha$ and $\lambda$ that leads to the minimal MSE was chosen. $\alpha$ and $\lambda$ values in $(\alpha, \lambda)$ pairs from each cross-validation run were then standardized so that $\alpha$ and $\lambda$ values are on the same scale and centered to the origin of the $\alpha - \lambda$ plane. $(\alpha_{min}, \lambda_{min})$ was then chosen as the $(\alpha, \lambda)$ point closest to the origin. With this $(\alpha_{min}, \lambda_{min})$ value, the model was then retrained on the entire dataset. All non-zero terms and the corresponding regression coefficients were reported.

## 3    Results

PCA was first applied to the dataset consisting of 18 demographic and health factors for 118 countries. Cumulative data variance that is explained jointly by the first *n* PCs is shown in Figure 1A (with *n* represented on the x-axis). In particular, Figure 1A shows the first PC alone already accounts for 45% of the variance, while the first 9 PCs (PC1 – PC9), which we retain in further analysis, explain more than 85% (precisely, 89%).

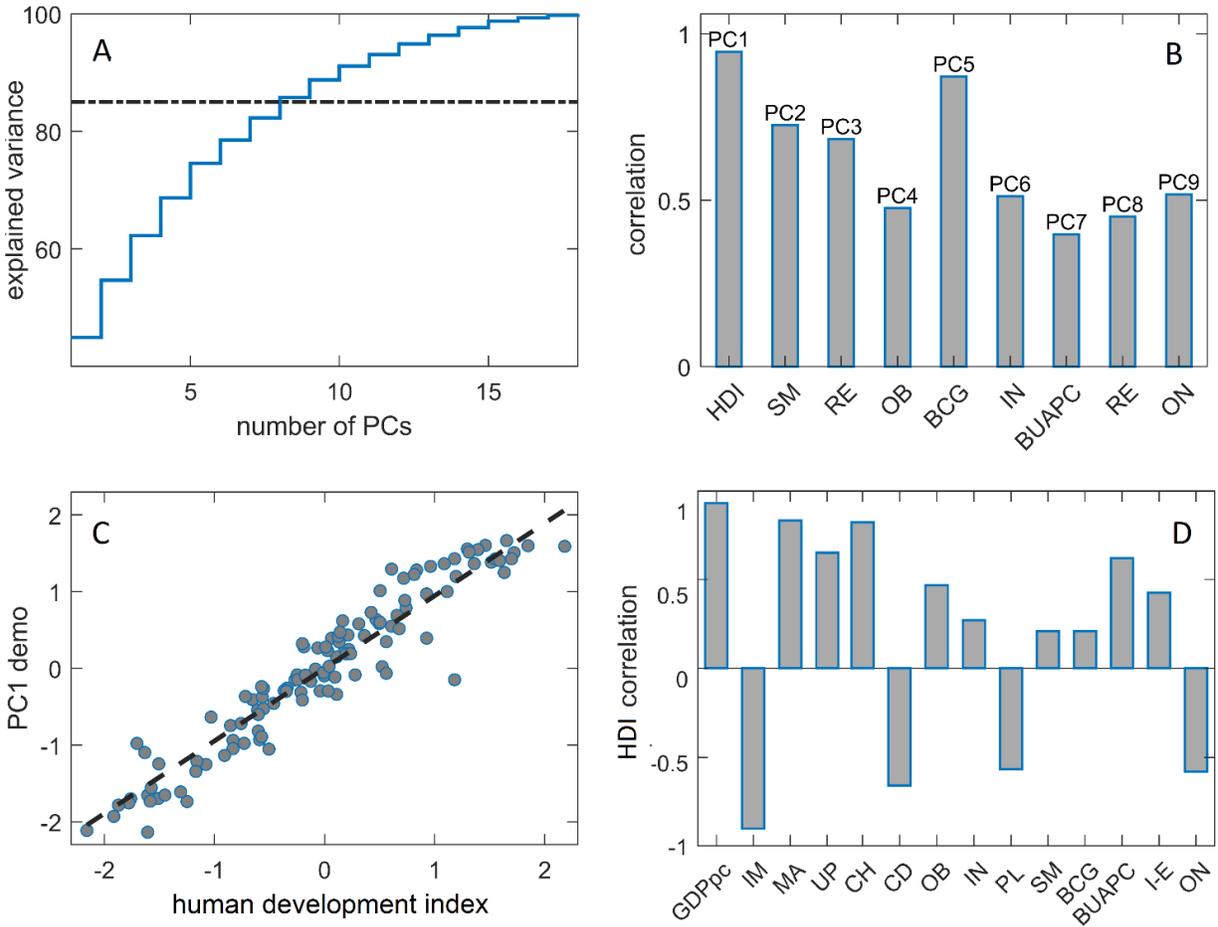

**Figure 1.** PCA for demographic data. **A)** Cumulative explained variance. **B)** Variables best correlated with demographic PCs. The labels above and below each bar present, respectively, the demographic PC and the variable with which that PC has the highest correlation. HDI – Human development index, SM – prevalence of smoking, RE – the percentage of refugees, OB – prevalence of obesity, BCG – BCG immunization coverage, IN – prevalence of insufficient physical activity, BUAPC – Built-up area per capita, ON – epidemic onset **C)** Scatter plot PC1 vs HDI. **D)** Correlations of selected demographic variables with HDI. GDPpc – GDP per capita, IM –infant mortality, MA – median age, UP – urban population, CH – average blood cholesterol level, CD – prevalence of chronic diseases, OB – prevalence of obesity, IN – prevalence of insufficient physical activity, PL – PM2.5 pollution, SM – prevalence of smoking, BCG – BCG immunization coverage, BUAPC – built-up area per capita, I-E – number of immigrants minus emigrants, ON – epidemic onset.

To obtain a basic interpretation of these nine PCs, we related each PC with the original (transformed) variable it is most correlated with. The corresponding associations – with the values of correlations coefficients presented on the y-axis – are shown in Figure 1B (however, one should have in mind that some PCs are highly correlated with more than one original variable, as we discuss in more detail below). Among all principal components, the PC1 and the PC5 have the highest correlation coefficients (close to 1) with individual demographic factors – the HDI and the BCG immunization coverage, respectively. Moderately high correlation coefficients (~0.75) characterize the relations between the PC2 and the prevalence of smokers, and the PC3 and the percentage of refugees, while the coefficient values of ~0.5 were obtained for the correlations of the PC4, the PC6, the PC7, the PC8 and the PC9 with, respectively, the prevalence of obesity, the prevalence of insufficient physical activity, the amount of the built-up area per person, the percentage of refugees, and the epidemic onset.

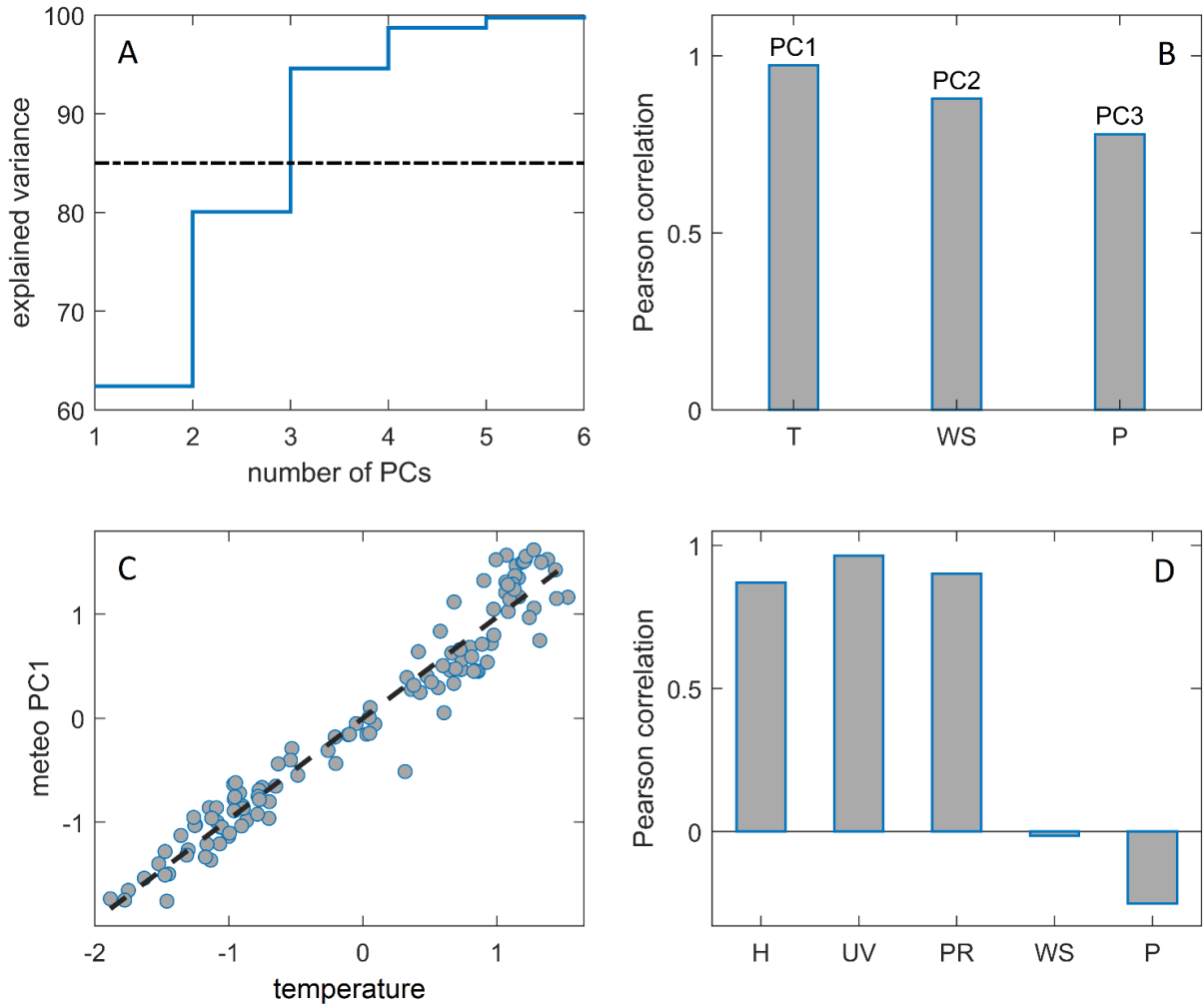

**Figure 2.** PCA for meteorological data. **A)** Cumulative explained variance. **B)** Variables best correlated with meteorological PCs. T – temperature, WS – wind speed, P- air pressure **C)** Scatter plot meteo PC1 vs temperature. **D)** Correlation of meteorological variables with temperature. H – specific humidity, UV – ultra-violet radiation index, PR – precipitation, WS – wind speed, P- pressure.

In particular, the first PC, accounting alone for the largest portion of the variance in the demographic data, is almost perfectly correlated with the Human Development Index (Fig. 1C). On the other hand, the HDI variable itself strongly correlates with several other demographic variables (Fig. 1D), most prominently with per capita GDP, infant mortality, and cholesterol levels. As elaborated in the Discussion section, such extremely high correlations will eventually preclude us from differentiating between the separate effects of each of these variables on $R_0$. On the other hand, the prevalence of obesity, the built-up area per person, and the epidemic onset are significantly correlated with the HDI (Fig. 1D), and thereby the PC1 (Fig. 1C), but they are markedly featured also in separate principal components (Fig. 1B), namely – the PCs 4, 7 and 9. This will help us to infer whether their specific (additional) contributions to the variance in the data (apart from that along the PC1) impact the virus transmissibility.

For meteorological factors for 118 countries was reduced similarly as for the demographic dataset. PCA generated six uncorrelated, orthogonal principal components. Thereby, the first PC alone explains 62% of the variance, while the first three PCs (PC1-PC3) capture 95%, which is significantly above the targeted 85% of the total variance (Fig. 2A). Pairwise correlations showed that the retained three PCs have the highest correlations with the temperature, the wind speed, and the air pressure, respectively (Fig. 2B), where the correlation of PC1 with the temperature is close to 1 (Figs. 2B and 2C). There are also notable correlations of the temperature with humidity, the levels of UV radiation, and precipitation (Fig. 2D). Therefore, PC1 presents seasonality, i.e., a set of mutually correlated meteorological variables that can be related to yearly weather changes. PCA, thus, effectively separated the impacts of seasonality (PC1), the wind speed (through the PC2), and the air pressure (through the PC3). The variables determining the PC1 are also correlated with the HDI. These inter-dataset correlations are not resolved at this level by our PCA and represent the trade-off that allows interpreting the PCs more easily within each of the two smaller, thematic groups of factors.

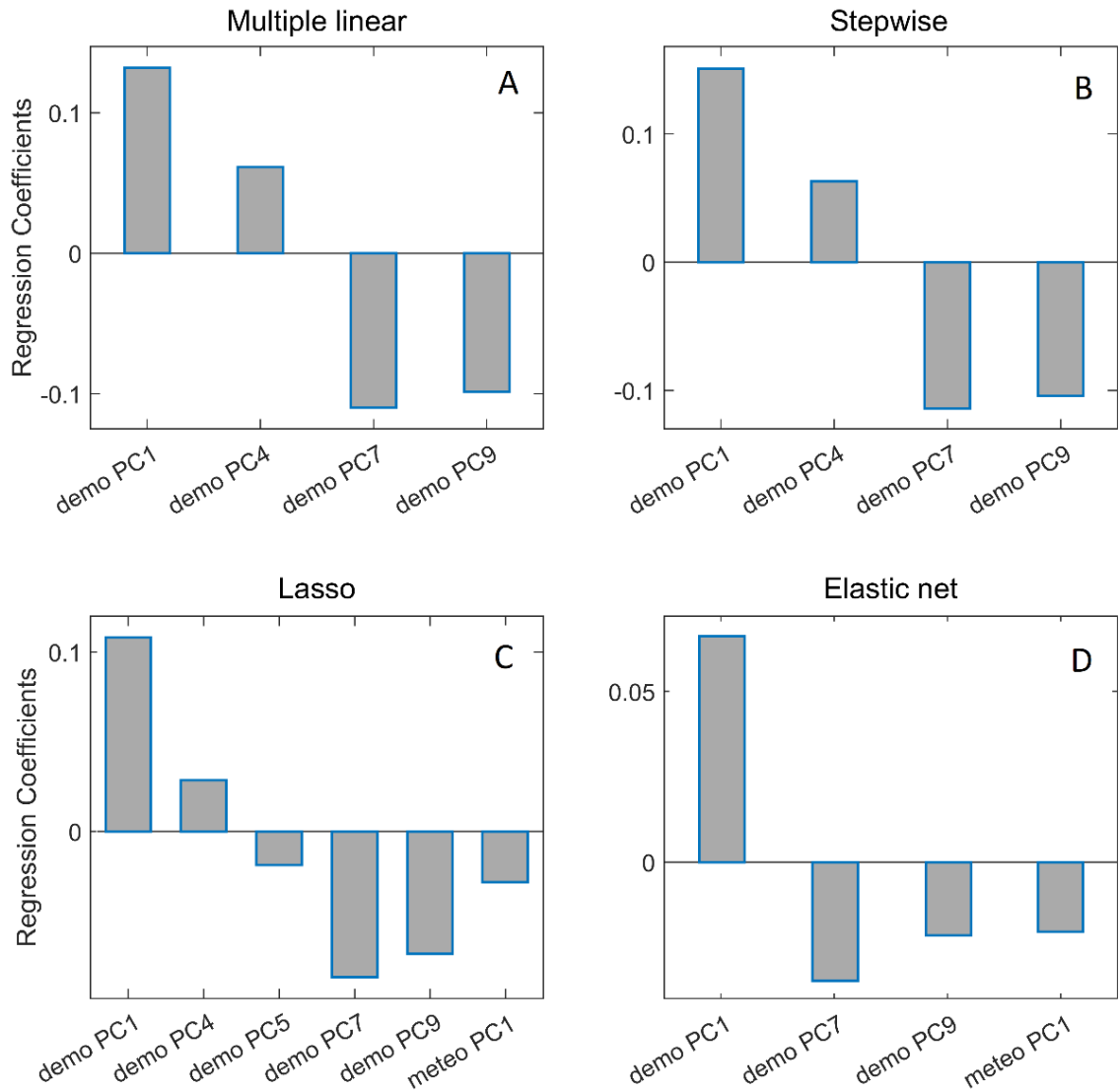

**Figure 3.** Results of: **A)** multiple linear regression ("custom") method, **B)** Stepwise regression, **C)** Lasso regression, and **D)** Elastic net regression. Bar charts represent the values of regression coefficients for each of the PCs selected by the method.

After PCA, we applied the linear regression analysis using four different methods, as explained in Methods. The first, "custom" method included the additional step of "preselecting", i.e., further narrowing down the number of PCs that will enter the final regression analysis. The multiple linear regression, applied on the group of 9 demographic PCs, selected 1st, 4th, 7th and 9th component as the most relevant predictors of $R_0$ (the remaining 5 PCs appeared in the linear regression with p values above 0.05 threshold and were consequently excluded from the further analysis, see Table 2. Analogously, the "preselection" of meteorological PCs singled out the 1st component as the only statistically relevant predictor of $R_0$ from this group (see Table 3). The multiple linear regression was then applied on these five selected PCs (four demographical and one meteorological) and yielded a regression model with the corresponding linear coefficients represented in Figure 3A (see also Table 4). Meteo PC1 component does not appear in the results of the custom method due to the lack of statistical significance ($p>0.305$) in the final regression. Thus, according to our custom regression methodology, weather parameters do not significantly influence $R_0$. $R_0$ in this model is therefore determined by a combination of demographic PC1, PC4, PC7, and PC9, where coefficients multiplying PC1 and PC4 are positive, while for PC7 and PC9 are negative. As can be inferred from the values represented in Fig 3A, the demographic PC1 has the most dominant influence on $R_0$ – a robustly obtained result throughout all four methods (see below).

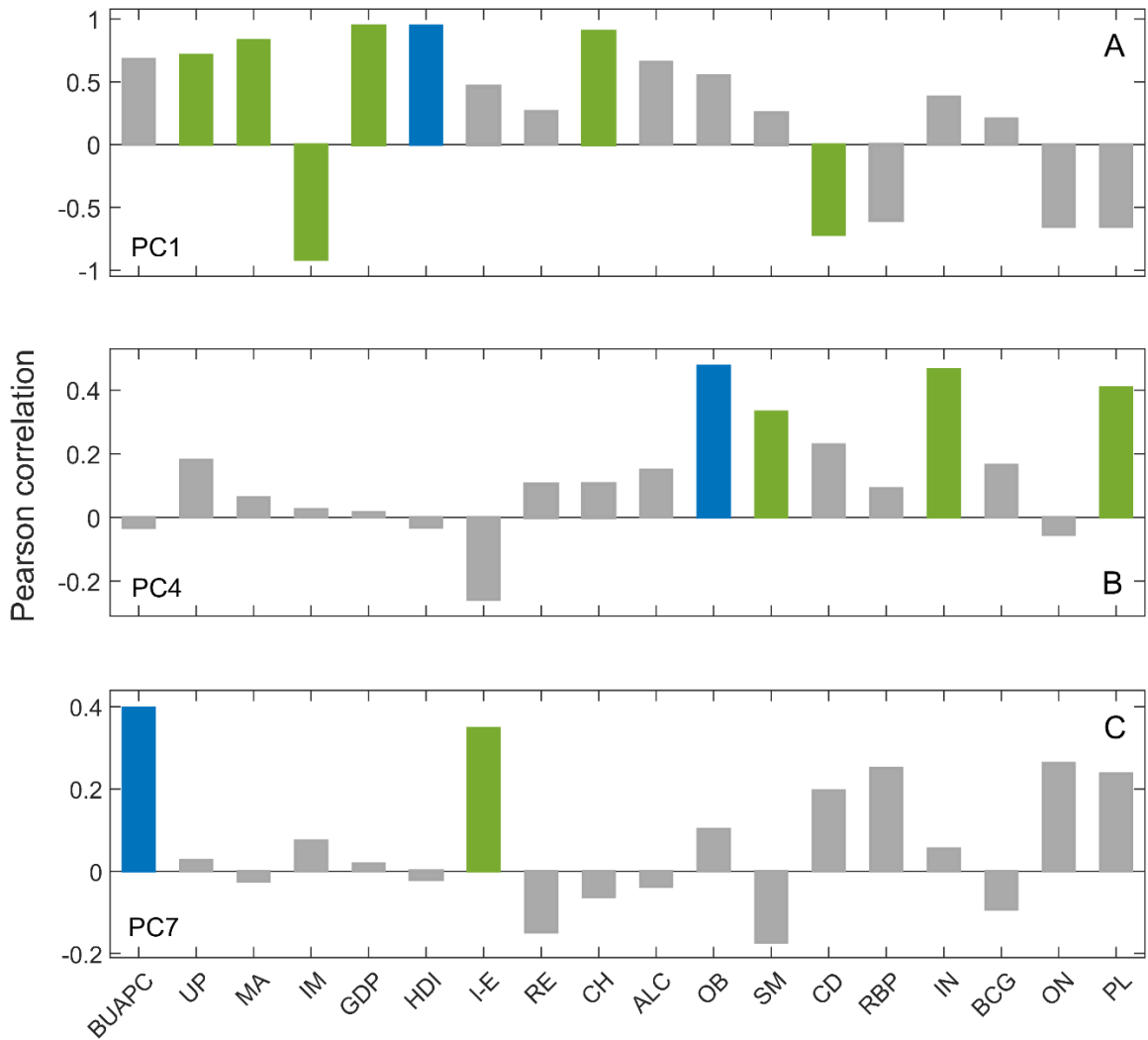

**Figure 4:** Pearson correlation coefficients between principal components and demographic variables for **A)** demo PC1, **B)** demo PC4, and **C)** demo PC7. BUAPC – built-up area per capita, UP – urban population, MA – median age, IM – infant mortality, GDP – gross domestic product per capita, HDI – human development index, I-E – number of immigrants minus emigrants, RE – percentage of refugees, CH – average blood cholesterol level, ALC – alcohol consumption per capita, OB – prevalence of obesity, SM – prevalence of smoking, CD – prevalence and severity chronic diseases, RBP – prevalence of raised blood pressure, IN – prevalence of insufficient physical activity, BCG – BCG immunization coverage, ON – epidemic onset, PL – long-term average PM2.5 pollution.

We have already related each of these four PCs with the dominantly correlated variable (Figure 1B), but a more detailed interpretation of the results is obtained if all significant correlations (not just the dominant one) are taken into account. In addition to the very high correlation with HDI, demographic PC1 is also highly positively correlated with GDP, cholesterol levels, median age, and percentage of the urban population, while it is highly negatively correlated with infant mortality and the prevalence of chronic diseases (Figure 4A). Such strong correlations with HDI, GDP, IM, MA, and UP show that this component indeed expresses an overall, both social and financial, the prosperity of the country (which seemingly also goes hand in hand with high average cholesterol levels and low prevalence of COVID-19 relevant chronic diseases). Similarly, by considering the correlations of demo PC4 with all demographic variables, we see that this component is significantly positively correlated not only with obesity but also with smoking, physical inactivity, and air pollution (Figure 4B) – in other words, with major indicators of an unhealthy lifestyle and living conditions. Apart from its correlation with the BUAPC parameter, the component demo PC7 is also significantly positively correlated with net migration (Figure 4C). In the case of the demo PC9 component, its only significant correlation is with the onset variable. Results of the custom method can therefore be summarized as follows: the country's prosperity, as well as unhealthy living conditions and lifestyle, tend to increase the value of $R_0$, while the larger built-up area per person and the later epidemic outbreak tend to slow the spread of the disease. Also, the results seem to indicate – via demo PC7 component – a surprising diminishing effect of the net migration on the rate of epidemic progress (though the sign of this variable may not be easy to interpret, as the net migration is a difference of two quantities).

Equivalently to Figure 3A, Figures 3B, 3C, and 3D represent the results of, respectively, Stepwise, Lasso, and Elastic net regression. Results (and the corresponding graph) of the Stepwise method almost coincide with the results of our custom method – in spite that in the Stepwise regression (as well as in Lasso and Elastic net methods), there is no intermediate "preselection" step.

Lasso results, shown in Figure 3C, find two additional PCs as relevant: demo PC5 and meteo PC1 (in addition to demo PC1, demo PC4, demo PC7, and demo PC9). The component demo PC5, appearing in Lasso results with a small negative coefficient, is significantly correlated only with the BCG variable, hinting at possible beneficial effects of BCG vaccination. Meteorological principal component meteo PC1 reflects seasonality (see above). In addition to supporting the conclusions of the custom and stepwise methods, the Lasso method thus also implicates a significance of seasonality changes and (to some extent) BCG vaccination in reducing the rate of SARS-CoV2 spread.

The results of the Elastic net method, shown in Figure 3D, are again a bit more restrictive. While further bolstering our confidence in the importance of demo PC1, demo PC7, and demo PC9, these results also reinforce that the seasonal weather variables influence the COVID-19 epidemic (in agreement with the Lasso method) but, for the first time, we do not find an indication of the relevance of the unhealthy lifestyle and living conditions – as revealed by the absence of demo PC4 component in Figure 3D.

Finally, as much as the PCs appearing in Figure 3 are important, the absence of the remaining PCs in the results can be of comparative significance for some of our conclusions. For example, we note that PCs highly correlated with the urban population, alcohol consumption, and chronic diseases do not show up as relevant in any of the methods used. While it is true that these variables are moderately correlated with demo PC1, absence in the results of additional PCs tied with these variables supports the view that these variables are not directly influencing $R_0$ value, but only via indirect relation to the country's prosperity.

## 4    Discussion

Our goal was to identify the most predictive factors influencing the risk of the SARS-CoV-2 virus spreading in a population in the absence of any epidemic mitigation measures. Since many potentially relevant factors strongly correlate with each other, we divided them into two groups – meteorological and sociodemographic – and applied the PCA to the variables in each group. In this way, we were able to decorrelate variables within each group while still retaining intuitive interpretation for the new variables (demographic and meteorological PCs) used in further analysis. While a similar approach was proposed in (Notari and Torrieri, 2020; Notari, 2021), dimensionality reduction and predictor decorrelation through PCA were, in our study, combined with different variable selection and regularization techniques – namely, Lasso and Elastic net - to select PCs that are most predictive of $R_0$ for COVID-19 epidemics. Examining correlations of these PCs with the original variables allowed pinpointing the main drivers of COVID-19 transmissibility. Therefore, this approach, which is reminiscent of the analysis of complex data in systems biology and bioinformatics, to our knowledge represents the most thorough effort to disentangle true COVID-19-related effects from indirect correlations.

Three principal components are robustly selected as the most important predictors by all the methods. Of these, the prosperity of the country has the most significant influence on $R_0$: the spread of the epidemic is faster in economically more developed countries. Specifically, this is the most dominant PC from the demographic group of variables, which is by far most important in explaining $R_0$, and very strongly correlated with HDI (Pearson's correlation coefficient $r$=0.95) and GDP ($r$= 0.94) – therefore effectively reflecting prosperity and wealth. The second PC is dominantly related to the built-up area per person (BUAPC), and the third with the epidemic onset, where the increase of these reduces the infection spread. We also robustly obtained (by three out of four methods) that unhealthy living conditions/lifestyle is another important factor that exacerbates the epidemic - this PC is dominantly (and consistently positively) correlated with obesity, physical inactivity, smoking, and air pollution. The group of four weather conditions, represented by seasonality, was selected by two independent methods. One is the Elastic net, which is well adapted for selecting among correlated variables (Zou and Hastie, 2005; Hastie et al., 2009) - note that correlations between meteorological and demographic PCs were not abolished by our approach. The PC dominantly correlated with BCG immunization appears only in Lasso regression. In favor of our results, we notice that the analysis of 13 selected variables conducted by Notari and Torrieri (2020) pointed to several similar factors as significant, such as the epidemic starting date, temperature, and the factors characteristic of general poor health and, thus, lower resistance to the virus infection (i.e., smoking, obesity and lung cancer as related to the chronic diseases).

HDI (alternatively, GDPpc) shows the highest correlation with the first demographic PC, which singles out this variable as the main index quantifying the virus transmissibility risk. Higher HDI leads to a higher rate of social contacts and more intense population mixing (especially over long distances), as high HDI is strongly related to high GDPpc implying intensive economic activity, trade, and transportation, including large-distance flights (Allel et al., 2020; Gangemi et al., 2020). While it is true that the intensity of social binding might often be even higher in low-income societies, such interpersonal relations are prevalently of spatially-local character and may less contribute to a large scale epidemic than the intense business activity (trade, transport, tourism, education), which connects and mixes people over large distances/areas. Thus, in our view, much higher contact frequency (especially of contacts that mix individuals over large distances) in societies with higher HDI is likely the main cause behind the dominant role of the first demographic PC in explaining $R_0$.

There is, however, no reason to expect this relation of HDI with $R_0$ to uniformly hold across the entire range of HDI (prosperity PC) values: indeed, analysis performed in (Notari and Torrieri, 2020) shows that GDP (strongly correlated to HDI and our prosperity PC) ceases to be a significant predictor of COVID-19 transmission rate once the low-income countries are excluded. One possible interpretation is that above a certain threshold of HDI, there is a focus on fully developed modern societies. The impact of further increasing country prosperity on the frequency of social contacts then becomes overshadowed by other factors. This was obtained in our recent study of the rate of COVID-19 transmission in USA states (Milicevic et al., 2021), where different factors, in the first place pollution, become dominant in that dataset.

This explanation of the impact of prosperity PC is still a hypothesis, which should be tested by further investigating the relation of HDI to some more direct measures of frequency of actual physical contact between individuals (unfortunately, such measures are not readily available) (Bontempi et al., 2020). We think that the obtained strong association of the PC closely related with GDP/HDI with $R_0$ is an important result (allowing a better understanding of epidemic risks), regardless of whether our interpretation (entirely) captures the true underlying causes of this connection.

An advantage of our approach is that it is based on the analysis of $R_0$ rather than other measures used as transmissibility proxies. The most commonly used measure, confirmed case counts, strongly depends on the number of performed tests, which is generally much higher in high-GDPpc countries, so the analysis would become strongly influenced by testing policies. For example, in (Allel et al., 2020), the importance of HDI for predicting cumulative case counts was noted. However, this perceived effect may be due to the lack of testing in lower-income countries, as already noted in (Notari, 2021), rather than genuine HDI influence. Our results are, on the other hand, insensitive to the testing capacity differences. Namely, since our $R_0$ estimation procedure relies on the scale-invariant slope of the case growth curve logarithm (in the distinct early exponential phase) (Djordjevic et al., 2021), it does not depend on the percentage of the infected individuals that get tested/diagnosed (i.e., multiplying daily case counts by arbitrary number will not affect $R_0$). As long as this percentage is roughly constant (i.e., testing is performed consistently) during the (relatively short) examined period, our $R_0$ estimate remains valid irrespectively of whether the country must reserve testing kits only for symptomatic cases, or the country has the means to (preventively) test asymptomatic individuals. On the other hand, if testing policies/criteria happen to irregularly change during the observed period (e.g. due to a sudden lack of testing resources), this would be accounted for (i.e., filtered out) as described in Methods. Therefore, our analysis indeed strongly suggests that HDI/GDPpc are the main/genuine predictors of COVID-19 spread in the population.

Many correlations previously reported between SARS-CoV-2 transmissibility and various weather, sociodemographic, and health factors [see e.g. (Li et al., 2020; Salom et al., 2021)] may be captured by HDI. From our results, one can note that several demographic factors significantly correlate with both HDI/GDPpc and the first demographic PC, but are not noticeably related with other demographic PCs (4,5,7,9) that significantly contribute to $R_0$. These demographic factors can be further divided into two groups using the correlation of BUAPC with HDI as the reference. The percentage of the urban population, the prevalence of alcohol consumption, and chronic diseases, which have similar (just somewhat higher) correlations with HDI compared to BUAPC, comprise the first group. Their absence from the independent PCs significantly related with $R_0$ (in contrast to BUAPC that prominently appears in the demographic PC7), indicates that they do not have independent effects on $R_0$. Consequently, their significant correlation with $R_0$ (Salom et al., 2021) is likely due to their generic correlation with HDI, rather than a consequence of the independent effect that they exhibit on $R_0$. This result is especially interesting for the percentage of the urban population, whose relation with $R_0$ is sometimes taken for granted (Carozzi, 2020). It also explains the previously obtained negative correlation of the prevalence of chronic diseases with $R_0$, where one might expect the opposite, as it is generally known that people with chronic diseases are seriously affected by COVID-19 (Zheng et al., 2020). We can now claim that this result is due to a generically lower incidence of chronic diseases in more developed countries (i.e., due to their significant negative correlation with HDI) rather than a direct effect on $R_0$.

A significant positive correlation with the first demographic PC is also exhibited by the net economic immigration (the difference between immigrants and emigrants), population median age, infant mortality, and the average blood cholesterol level. However, their correlation with HDI is very high (in distinction to the three factors mentioned before), i.e., visibly higher compared to the correlation of BUAPC with HDI. So, even though they do not appear in demographic PCs that significantly contribute to $R_0$ other than PC1, we cannot make any reliable conclusion about their direct effect on $R_0$ based on our analysis. Therefore, it is relevant to discuss evidence from other sources, i.e., possible mechanisms for distinguishing their direct influence on $R_0$. Regarding infant mortality, a mechanism of its direct contribution to $R_0$ is hard to imagine, so its involvement in PC1, and high negative correlation with $R_0$, is almost certainly an indirect consequence of this variable being a proxy of HDI (Ruiz et al., 2015). On the other hand, the median age and the blood cholesterol level are real contenders for direct $R_0$ modifiers, as mechanisms for their contribution to COVID-19 transmissibility have been proposed. Aging is generally associated with the weakening of the immune response to infectious diseases making the elderly more susceptible to the viruses like the SARS-CoV-2 (Pawelec and Larbi, 2008). Additionally, many of them due to some chronic diseases take ACE inhibitors and angiotensin-receptor blockers which cause an increased expression of ACE2 serving as a receptor for the SARS-CoV-2 virus entry (Shahid et al., 2020). Their residing in care homes, which is particularly common in high-income countries, also well suits the spreading of the infection (Kapitsinis, 2020). Similarly, high cholesterol levels can increase susceptibility to the infection by SARS-CoV-2 through systemic adverse effects on the immune and inflammatory responses and through direct implication in the virus life cycle (especially at the level of its endocytosis). To that end, statins, blocking cholesterol synthesis, were proposed for usage in COVID-19 treatment, which is supported by studies showing that previous statin usage is associated with a milder pneumonia outcome in the case of several other viral infections (Frost et al., 2007; Schmidt et al., 2020).

Other demographic PCs (4,5,7,9) that are selected to have a significant influence on $R_0$ are by construction independent (decorrelated) from PC1. Variables associated with these PCs can be interpreted as effects on $R_0$ independent from those related to PC1. These variables then importantly identify corrections to the main effect of HDI/GDPpc. Specifically, these are indoor area available to an individual and the net immigration (demographic PC7), the delay in the epidemic onset, which is associated with more awareness of the virus threat (demographic PC9), the prevalence of unhealthy lifestyle and environment (demographic PC4), and the weather seasonality (meteorological PC1). The slower spread of the virus with a larger built-up area per capita, as an independent and significant $R_0$ predictor, is an interesting and new result, though intuitively plausible. It can be understood as having a less crowded indoor space (where the virus transmission dominantly happens) so that people are less exposed to each other and the virus. For example, on the Diamond Princess cruise ship, both the population density and $R_0$ were estimated as four times greater than those in Wuhan (Rocklöv and Sjödin, 2020). On the other hand, a correlation of the virus transmissibility with the large territory population density is weakly established in the literature, whereby it seems that one should rather seek a correlation with a local population density, directly determining the number of contacts that an individual can make (Garland et al., 2020; Diao et al., 2021).

A positive contribution to the transmissibility is also made by the principal component strongly correlated with the onset variable, representing the number of days from February 15th to the epidemic's start in a particular country. The importance of the delay in the epidemic onset may be due to the psychological effect of hearing the news about the spread of COVID-19 in other countries (Khajanchi et al., 2020). Namely, the longer the epidemic was growing outside of a particular country, the larger impact this had on its people to change their usual behavior to prevent the infection, which could slow down the virus transmission even before the introduction of the official intervention measures (Salom et al., 2021).

Another distinguished principal component appears to encompass multiple indicators of an unhealthy lifestyle and environment – specifically, the prevalence of obesity, physical inactivity, and smoking, together with the level of air pollution. We obtained that all these factors promote virus transmission. It is well established that they can impair immune function and adversely affect different organ systems. Furthermore, their association with mechanisms specifically facilitating the infection by the SARS-CoV-2 virus has been proposed (Domingo and Rovira, 2020; Heidari-Beni and Kelishadi, 2020; Haddad et al., 2021). Notably, the association of air pollution with COVID-19 transmission has been shown in several studies (Bashir et al., 2020; Coccia, 2020b; Milicevic et al., 2021). Two of the remaining relevant PCs are strongly determined by temperature (and/or three other highly related weather factors) and the prevalence of BCG vaccinated children, respectively. Although not selected by all the methods, the weather component seems important as it was chosen by the Elastic net algorithm (in addition to Lasso), which is specifically designed to deal with (highly) correlated variables, and yet it did not exclude this PC despite its correlation with the first demographic PC. Moreover, a decrease of the transmissibility with the temperature increase appears as a robust result in COVID-19 literature (Haque and Rahman, 2020; Rosario et al., 2020; Sarkodie and Owusu, 2020), although conflicting conclusions are also present (Xie and Zhu, 2020; Islam et al., 2021; Srivastava, 2021). Higher temperatures may shorten the period of virus viability in aerosols, enhance the immune system functioning, and/or impact the time that people spend together in poorly ventilated indoor spaces (Notari, 2021). Since temperature is highly positively correlated with the intensity of UV radiation, humidity, and the level of precipitation, we cannot exclude the possibility that some of these other factors are in a significant causal relationship with virus transmissibility. Importantly, some experimental findings support the inactivating effects of high temperature, humidity, and UV radiation on SARS-CoV-2 and related viruses (Casanova et al., 2010; Chan et al., 2011; Heiligloh et al., 2020; Sagripanti and Lytle, 2020; van Doremalen et al., 2020). Anyhow, our results suggest the dependence of virus transmissibility on seasonal weather variations related to our meteorological PC1 component. On the other hand, while some authors suggest the importance of wind speed in virus dissemination (Coccia, 2020b; Coccia, 2020a; Sarkodie and Owusu, 2020; Islam et al., 2021), our study could not confirm this connection – as reflected by the absence of the wind-related principal component meteo PC2 in our results. This might be a consequence of a strong interplay between wind speed and pollution effects (Coccia, 2021b) that our study is not suited to detect. The last demographic principal component occurred as important only in Lasso regression. It, however, closely follows the extent of BCG vaccination, which is known to provide some protection against various respiratory tract infections through the induction of the trained immunity (O'Neill and Netea, 2020), so BCG immunization may influence the SARS-CoV-2 spread, although, according to our results, to a lesser extent than the other discussed factors.

Our study is also an example of how assessing the effect of one factor while controlling for the presence of other relevant variables can change the obtained conclusions. We will illustrate this with four examples, where we obtained qualitatively different conclusions, compared to single-variable correlation analysis (Salom et al., 2021): built-up area per capita (BUAPC), net migration, air pollution, and raised blood pressure. BUAPC showed an absence of a significant correlation with $R_0$ (Salom et al., 2021), which is a consequence of canceling the two effects. The first is its direct effect on $R_0$ (exhibited through demographic PC7), which tends to slow the spread of COVID-19 in a population. The second is collinearity with PC1, which reflects a generic correlation of BUAPC with GDPpc, caused by more construction (higher built-up area) per capita with the increase in GDPpc. Our combination of PC and regression analysis revealed this non-trivial conclusion, which cannot (even qualitatively) be obtained from the pairwise correlation analysis.

Similar reasoning, though perhaps harder to understand intuitively, applies to net migration. Net migration is also significantly positively correlated with HDI (and consequently also with PC1), reflecting a generic tendency of immigrants to flow to countries with higher GDPpc. The direct effect of net immigration (exhibited through PC7) is harder to intuitively understand, as I-E negatively contributes to $R_0$, so that faster spread (at least in the initial phase of the epidemic) appears to be associated with a higher number of emigrants. As these are economic migrations (to be distinguished from the movement of refugees), possibly the part of the emigrants returned to their countries with the pandemic's start. The significant effect of net immigration on $R_0$ (inferred through our analysis) is again highly non-trivial and in the opposite direction from the positive pairwise correlation of $R_0$ with I-E. Refuges (i.e., percentage of refugee population by country) exhibit high correlations only with PC3 and PC8, where neither significantly contributes to $R_0$. There is also no significant pairwise correlation of refugees with $R_0$, which robustly shows that this variable does not affect transmissibility.

Pollution negatively contributes to demographic PC1 (with the corresponding negative correlation with HDI) and positively to demographic PC4. The pairwise correlation between the pollution and $R_0$ is negative (-0.31), which is counterintuitive, as it is generally expected that higher pollution should increase COVID-19 transmissibility. This is, however, an artifact of negative correlation between pollution and HDI, while its genuine (direct) effect on $R_0$ is reflected through PC4. Our analysis, therefore, revealed the direct effect of long-term air pollution on transmissibility, which is consistent with previously published observations that it can damage the respiratory system and reduce resistance to infections (Domingo and Rovira, 2020; Fattorini and Regoli, 2020) but opposite to naive pairwise correlation analysis.

Finally, raised blood pressure also shows a statistically significant (counterintuitively negative) correlation with $R_0$. However, in addition to PC1, raised blood pressure shows a notable correlation only with PC2, which does not significantly affect $R_0$. This indicates that the negative correlation of this variable with $R_0$ is a consequence of its generically negative correlation with HDI, instead of a direct effect on COVID-19 transmissibility.

## 5    Conclusion and Outlook

Numerous studies tried to assess the correlations of different factors with the SARS-CoV-2 virus transmissibility (Li et al., 2020; Notari and Torrieri, 2020; Salom et al., 2021), but the next step should be predicting the environmental risk of the high spreadability in a certain population (Allel et al., 2020; Coccia, 2020d; Gupta and Gharehgozli, 2020). Specifically, a relatively small number of the most influential meteorological and demographic factors should be selected for a predictive risk measure that is accurate enough and practical for use. Such risk assessment is very useful in guiding the future strategies of imposing epidemic mitigation measures (Coccia, 2021a).

We here demonstrated that taking into account joint effects of different factors can point to qualitatively different conclusions about their influence on the virus transmissibility than considering them individually (as in (Salom et al., 2021)). Utilizing a combination of PCA and feature selection techniques, we were able to disentangle with high confidence which variables independently (and significantly) influence the rate of the infection spread, and which have an only indirect influence or no influence at all (here found for alcohol consumption, chronic diseases, percentage of the urban population, raised blood pressure and refugees).

While PCA brings clear advantages to regression analysis, such as working with a smaller number of variables and abolishing collinearity, the main disadvantage is harder interpretation in terms of original variables. We were, however, able to interpret PCs that significantly affect $R_0$, so that the main driving factors behind COVID-19 transmissibility are i) the country's wealth/development level, corrected by the available indoor space per person and net immigration), ii) pollution levels and some of the unhealthy living factors, iii) spontaneous behavior change due to developing epidemics, iv) weather seasonality and v) possibly (marginally) BCG vaccination. These conclusions, and the direction of the corresponding effects, crucially depend on the more complex analysis performed here.

Certain limitations of this study should also be addressed. When the alignment between certain variables is too high, even the analysis performed here cannot differentiate between the factors genuinely affecting $R_0$ and mere accidental correlations. In such cases, further, specifically designed (such as targeted epidemiological) studies are needed. For example, based on this analysis alone and due to the very high correlation between the cholesterol levels and HDI/GDP, it cannot be excluded that cholesterol is a contributing factor to the observed significance of the PC1 component, in addition to the country's prosperity that likely mimics the contact rate in population (as a crucial disease transmission property). For this reason, our research suggests that a separate study of cholesterol levels in the COVID-19 context (e.g., by measuring cholesterol blood levels along with PCR tests) could be, potentially, of high value since a hypothetical unexpected discovery of inherent cholesterol importance could lead to novel treatments of SARS-CoV-2 infection. Similarly, studies that disentangle the effect of the overall country's prosperity from the intrinsic effects of median age on $R_0$ would be also quite welcome. Furthermore, some of the variables that were not included in our study, but were found as significant by other studies (Notari and Torrieri, 2020), and are closely related to HDI, such as lung cancer prevalence (Youlden et al., 2008), or frequency of tourist arrivals (Anggraeni, 2017) may also contribute to the strong association of HDI with $R_0$, which we observe. Another problem might occur due to possible untrivial interactions between different factors: for example, the effects of pollution on the airborne COVID-19 transmission can be significantly dependent on the wind speed (Coccia, 2020b; 2021b). Unfortunately, our study is not well-suited to detect such effects (while, in principle, these effects could be included through higher-order terms in regression, this might lead to overfitting). A non-trivial future task may be a systematic attempt to unify (and standardize, to the extent possible) different approaches and variables used in the studies of the environmental factor effects on COVID-19 transmissibility.

Some suggestions that could improve COVID-19 mitigation policies may already follow from this study. Deliberately reducing (in the long run) the overall population prosperity, which we identified as the most important predictor of $R_0$, in an attempt to indirectly reduce the frequency of social contacts is impractical. However, the widely implemented lockdowns, proven to inhibit the infection spread, are the measures that reduce the frequency of interpersonal contacts, but also economic activity, so the epidemic behavior seems consistent with the predicted HDI dependence. Our conclusions about the importance of HDI, as a predictor of $R_0$, could be further tested by studies of epidemiological relevance of higher resolution HDI-analogs, such as Subnational HDI (SHDI) or City Development Index (CDI). If HDI and GDP parameters are confirmed to dominantly influence $R_0$ values simply since they highly and naturally correlate with the frequency of social contacts (as we anticipate to be the case), identifying this as one of the major factors is not without implications. That is, recognizing the importance of this parameter can help make better predictions of the disease dynamic and locate in advance high-risk spots/areas.

Moreover, since we identified the built-up area per person as another relevant factor, future urban policies should put more emphasis on avoiding very high population concentrations (Coccia, 2020c). The relevance of the onset variable may indicate the importance of timely informing the public about all the risks and perils of the incoming epidemic. More obviously, pointing to unhealthy living conditions and lifestyle as a significant driver of COVID-19 pandemic, underscores the importance of measures to reduce obesity, smoking prevalence, physical inactivity, and air pollution (Coccia, 2021a), also by subsidizing energy from renewable resources (Coccia, 2020c). Overall, the results presented here emphasize the importance of understanding environmental factors that impact transmissibility for both COVID-19 and other (including potential future) infectious diseases.

## 6    Conflict of Interest

The authors declare that the research was conducted in the absence of any commercial or financial relationships that could be construed as a potential conflict of interest.

## 7    Author Contributions

MarD, IS, MagD and OM conceived the research. The work was supervised by MarD and IS. Code writing and data analysis by MarD and SM, with help of MagD and OM. Figures and tables made by SM with the help of MagD. A literature search by AR. The manuscript was written by IS, AR, and MarD, with help of MagD.

## 8    Data Availability Statement

Data is available through Salom et al., 2021.

## 9    Funding

This work was partially supported by the Ministry of Education, Science and Technological Development of the Republic of Serbia.

## 10    References

Alexopoulos, E.C. (2010). Introduction to multivariate regression analysis. *Hippokratia* 14(Suppl 1):23-28.

Allel, K., Tapia-Muñoz, T., and Morris, W. (2020). Country-level factors associated with the early spread of COVID-19 cases at 5, 10 and 15 days since the onset. *Glob. Public Health* 15(11):1589-1602. doi: 10.1080/17441692.2020.1814835.

Anggraeni, G.N. (2017). The relationship between numbers of international tourist arrivals and economic growth in the Asean-8: Panel data approach. *JDE (Journal of Developing Economies)* 2(1):40-49. doi: 10.20473/jde.v2i1.5118.

Bashir, M.F., Ma, B.J., Bilal, Komal, B., Bashir, M.A., Farooq, T.H., et al. (2020). Correlation between environmental pollution indicators and COVID-19 pandemic: A brief study in Californian context. *Environ. Res.* 187:109652. doi: 10.1016/j.envres.2020.109652.

Bontempi, E., Vergalli, S., and Squazzoni, F. (2020). Understanding COVID-19 diffusion requires an interdisciplinary, multi-dimensional approach. *Environ. Res.* 188:109814. doi: 10.1016/j.envres.2020.109814.

Carozzi, F. (2020). Urban density and COVID-19. *Institute for the Study of Labor (IZA)* [Online], 13440. Available at: https://ssrn.com/abstract=3643204.

Casanova, L.M., Jeon, S., Rutala, W.A., Weber, D.J., and Sobsey, M.D. (2010). Effects of air temperature and relative humidity on coronavirus survival on surfaces. *Appl. Environ. Microbiol.* 76(9):2712-2717. doi: 10.1128/AEM.02291-09.

Chan, K.-H., Peiris, J.M., Lam, S., Poon, L., Yuen, K., and Seto, W.H. (2011). The effects of temperature and relative humidity on the viability of the SARS coronavirus. *Adv. Virol.* 2011:734690. doi: 10.1155/2011/734690.

Coccia, M. (2020a). The effects of atmospheric stability with low wind speed and of air pollution on the accelerated transmission dynamics of COVID-19. *Int. J. Environ. Stud.* 78(1):1-27. doi: 10.1080/00207233.2020.1802937.

Coccia, M. (2020b). Factors determining the diffusion of COVID-19 and suggested strategy to prevent future accelerated viral infectivity similar to COVID. *Sci. Total Environ.* 729:138474. doi: 10.1016/j.scitotenv.2020.138474.

Coccia, M. (2020c). How (Un)sustainable Environments Are Related to the Diffusion of COVID-19: The Relation between Coronavirus Disease 2019, Air Pollution, Wind Resource and Energy. *Sustainability* 12(22):9709. doi: 10.3390/su12229709.

Coccia, M. (2020d). An index to quantify environmental risk of exposure to future epidemics of the COVID-19 and similar viral agents: Theory and practice. *Environ. Res.* 191:110155. doi: 10.1016/j.envres.2020.110155.

Coccia, M. (2021a). Effects of the spread of COVID-19 on public health of polluted cities: results of the first wave for explaining the dejà vu in the second wave of COVID-19 pandemic and epidemics of future vital agents. *Environ. Sci. Pollut. Res. Int.* 28(15):19147-19154. doi: 10.1007/s11356-020-11662-7.

Coccia, M. (2021b). How do low wind speeds and high levels of air pollution support the spread of COVID-19? *Atmos. Pollut. Res.* 12(1):437-445. doi: 10.1016/j.apr.2020.10.002.

Diao, Y., Kodera, S., Anzai, D., Gomez-Tames, J., Rashed, E.A., and Hirata, A. (2021). Influence of population density, temperature, and absolute humidity on spread and decay durations of COVID-19: A comparative study of scenarios in China, England, Germany, and Japan. *One Health* 12:100203. doi: 10.1016/j.onehlt.2020.100203.

Djordjevic, M., Djordjevic, M., Ilic, B., Stojku, S., and Salom, I. (2021). Understanding Infection Progression under Strong Control Measures through Universal COVID-19 Growth Signatures. *Global Challenges* 2021:2000101. doi: 10.1002/gch2.202000101.

Domingo, J., and Rovira, J. (2020). Effects of air pollutants on the transmission and severity of respiratory viral infections. *Environ. Res.* 187:109650. doi: 10.1016/j.envres.2020.109650.

Fattorini, D., and Regoli, F. (2020). Role of the chronic air pollution levels in the Covid-19 outbreak risk in Italy. *Environ. Pollut.* 264:114732. doi: 10.1016/j.envpol.2020.114732.

Frost, F., Petersen, H., Tollestrup, K., and Skipper, B. (2007). Influenza and COPD mortality protection as pleiotropic, dose-dependent effects of statins. *Chest* 131(4):1006-1012. doi: 10.1378/chest.06-1997.

Gangemi, S., Billeci, L., and Tonacci, A. (2020). Rich at risk: socio-economic drivers of COVID-19 pandemic spread. *Clin. Mol. Allergy* 18:12. doi: 10.1186/s12948-020-00127-4.

Garland, P., Babbitt, D., Bondarenko, M., Sorichetta, A., Tatem, A.J., and Johnson, O. (2020). The COVID-19 pandemic as experienced by the individual. arXiv [Preprint]. Available at: https://ui.adsabs.harvard.edu/abs/2020arXiv200501167G/abstract (Accessed March 12, 2021).

Gupta, A., and Gharehgozli, A. (2020). Developing a Machine Learning Framework to Determine the Spread of COVID-19. SSRN [Preprint]. Available at: https://papers.ssrn.com/sol3/papers.cfm?abstract_id=3635211 (Accessed March 12, 2021).

Haddad, C., Bou Malhab, S., Sacre, H., and Salameh, P. (2021). Smoking and COVID-19: A Scoping Review. *Tob. Use Insights* 14:1179173X21994612. doi: 10.1177/1179173X21994612.

Haque, S.E., and Rahman, M. (2020). Association between temperature, humidity, and COVID-19 outbreaks in Bangladesh. *Environ. Sci. Policy* 114:253-255. doi: 10.1016/j.envsci.2020.08.012.

Hassan, M., Bhuiyan, M., Tareq, F., Bodrud-Doza, M., Tanu, S., and Rabbani, K. (2021). Relationship between COVID-19 infection rates and air pollution, geo-meteorological, and social parameters. *Environ. Monit. Assess.* 193(1):29. doi: 10.1007/s10661-020-08810-4.

Hastie, T., Tibshirani, R., and Friedman, J. (2009). *The Elements of Statistical Learning: Data Mining, Inference, and Prediction.* New York: Springer.

Heidari-Beni, M., and Kelishadi, R. (2020). Reciprocal impacts of obesity and coronavirus disease 2019. *J. Res. Med. Sci.* 25(1):110. doi: 10.4103/jrms.JRMS_416_20.

Heilingloh, C.S., Aufderhorst, U.W., Schipper, L., Dittmer, U., Witzke, O., Yang, D., et al. (2020). Susceptibility of SARS-CoV-2 to UV irradiation. *Am. J. Infect. Control* 48(10):1273-1275. doi: 10.1016/j.ajic.2020.07.031.

Islam, N., Bukhari, Q., Jameel, Y., Shabnam, S., Erzurumluoglu, A.M., Siddique, M.A., et al. (2021). COVID-19 and climatic factors: A global analysis. *Environ. Res.* 193:110355. doi: 10.1016/j.envres.2020.110355.

Jolliffe, I.T. (2002). *Principal Component Analysis.* New York: Springer.

Kapitsinis, N. (2020). The underlying factors of the COVID-19 spatially uneven spread. Initial evidence from regions in nine EU countries. *Regional Science Policy & Practice* 12(6):1027-1045. doi: 10.1111/rsp3.12340.

Khajanchi, S., Sarkar, K., Mondal, J., and Perc, M. (2020). Dynamics of the COVID-19 pandemic in India. arXiv [Preprint]. Available at: https://arxiv.org/abs/2005.06286 (Accessed March 12, 2021).

Li, M., Zhang, Z., Cao, W., Liu, Y., Du, B., Chen, C., et al. (2020). Identifying novel factors associated with COVID-19 transmission and fatality using the machine learning approach. *Sci. Total Environ.* 764:142810. doi: 10.1016/j.scitotenv.2020.142810.

Lin, S., Wei, D., Sun, Y., Chen, K., Yang, L., Liu, B., et al. (2020). Region-specific air pollutants and meteorological parameters influence COVID-19: A study from mainland China. *Ecotoxicol. Environ. Saf.* 204:111035. doi: 10.1016/j.ecoenv.2020.111035.

Mallapaty, S. (2021). *What's the risk of dying from a fast-spreading COVID-19 variant?* [Online]. Available at: https://www.nature.com/articles/d41586-021-00299-2 [Accessed March 12, 2021].

Milicevic, O., Salom, I., Rodic, A., Markovic, S., Tumbas, M., Zigic, D., et al. (2021). PM2.5 as a major predictor of COVID-19 basic reproduction number in the USA.Environmental Research 201:111526.

Notari, A. (2021). Temperature dependence of COVID-19 transmission. *Sci. Total Environ.* 763:144390. doi: 10.1016/j.scitotenv.2020.144390.

Notari, A., and Torrieri, G. (2020). COVID-19 transmission risk factors. Authorea [Preprint]. Available at: https://authorea.com/users/360787/articles/482318-covid-19-transmission-risk-factors (Accessed March 12, 2021).

O'Neill, L.A., and Netea, M.G. (2020). BCG-induced trained immunity: can it offer protection against COVID-19? *Nat. Rev. Immunol.* 20(6):335-337. doi: 10.1038/s41577-020-0337-y.

Pawelec, G., and Larbi, A. (2008). Immunity and ageing in man: Annual Review 2006/2007. *Exp. Gerontol.* 43(1):34-38. doi: 10.1016/j.exger.2007.09.009.

Pope, P., and Webster, J. (1972). The use of an F-statistic in stepwise regression procedures. *Technometrics* 14(2):327-340. doi: 10.1080/00401706.1972.10488919.

Rahimi, N.R., Fouladi-Fard, R., Aali, R., Shahryari, A., Rezaali, M., Ghafouri, Y., et al. (2021). Bidirectional association between COVID-19 and the environment: A systematic review. *Environ. Res.* 194:110692. doi: 10.1016/j.envres.2020.110692.

Ran, J., Zhao, S., Han, L., Qiu, Y., Cao, P., Yang, Z., et al. (2020). Effects of particulate matter exposure on the transmissibility and case fatality rate of COVID-19: A Nationwide Ecological Study in China. *J. Travel Med.* 27(6):taaa133. doi: 10.1093/jtm/taaa133.

Rocklöv, J., and Sjödin, H. (2020). High population densities catalyse the spread of COVID-19. *J. Travel Med.* 27(3):taaa038. doi: 10.1093/jtm/taaa038.

Rosario, D.K.A., Mutz, Y.S., Bernardes, P.C., and Conte-Junior, C.A. (2020). Relationship between COVID-19 and weather: Case study in a tropical country. *Int. J. Hyg. Environ. Health* 229:113587. doi: 10.1016/j.ijheh.2020.113587.

Ruiz, J.I., Nuhu, K., McDaniel, J.T., Popoff, F., Izcovich, A., and Criniti, J.M. (2015). Inequality as a powerful predictor of infant and maternal mortality around the world. *PLoS One* 10(10):e0140796. doi: 10.1371/journal.pone.0140796.

Sagripanti, J.L., and Lytle, C.D. (2020). Estimated Inactivation of Coronaviruses by Solar Radiation With Special Reference to COVID-19. *Photochem. Photobiol.* 96(4):731-737. doi: 10.1111/php.13293.

Salom, I., Rodic, A., Milicevic, O., Zigic, D., Djordjevic, M., and Djordjevic, M. (2021). Effects of Demographic and Weather Parameters on COVID-19 Basic Reproduction Number. *Front. Ecol. Evol.* 8(524):617841. doi: 10.3389/fevo.2020.617841.

Sarkodie, S.A., and Owusu, P.A. (2020). Impact of meteorological factors on COVID-19 pandemic: Evidence from top 20 countries with confirmed cases. *Environ. Res.* 191:110101. doi: 10.1016/j.envres.2020.110101.

Schmidt, N., Wing, P., McKeating, J., and Maini, M. (2020). Cholesterol-modifying drugs in COVID-19. *Oxf. Open Immunol.* 1(1):iqaa001. doi: 10.1093/oxfimm/iqaa001.

Shahid, Z., Kalayanamitra, R., McClafferty, B., Kepko, D., Ramgobin, D., Patel, R., et al. (2020). COVID-19 and Older Adults: What We Know. *J. Am. Geriatr. Soc.* 68(5):926-929. doi: 10.1111/jgs.16472.

Srivastava, A. (2021). COVID-19 and air pollution and meteorology-an intricate relationship: A review. *Chemosphere* 263:128297. doi: 10.1016/j.chemosphere.2020.128297.

Tegally, H., Wilkinson, E., Giovanetti, M., Iranzadeh, A., Fonseca, V., Giandhari, J., et al. (2020). Emergence and rapid spread of a new severe acute respiratory syndrome-related coronavirus 2 (SARS-CoV-2) lineage with multiple spike mutations in South Africa. medRxiv [Preprint]. Available at: https://www.medrxiv.org/content/10.1101/2020.12.21.20248640v1 (Accessed March 12, 2021).

Tibshirani, R. (1996). Regression shrinkage and selection via the lasso. *J. Roy. Stat. Soc. B Met.* 58(1):267-288. doi: 10.1111/j.2517-6161.1996.tb02080.x.

van Doremalen, N., Bushmaker, T., Morris, D.H., Holbrook, M.G., Gamble, A., Williamson, B.N., et al. (2020). Aerosol and Surface Stability of SARS-CoV-2 as Compared with SARS-CoV-1. *N. Engl. J. Med.* 382(16):1564-1567. doi: 10.1056/NEJMc2004973.

World Bank (2020a). *GDP per capita (current US$)* [Online]. Available at: https://data.worldbank.org/indicator/NY.GDP.PCAP.CD [Accessed January, 2021].

World Bank (2020b). *World Bank Open Data* [Online]. Available at: https://www.worldbank.org/ [Accessed May, 2020].

Xie, J., and Zhu, Y. (2020). Association between ambient temperature and COVID-19 infection in 122 cities from China. *Sci. Total Environ.* 724:138201. doi: 10.1016/j.scitotenv.2020.138201.

Xie, Z., Qin, Y., Li, Y., Shen, W., Zheng, Z., and Liu, S. (2020). Spatial and temporal differentiation of COVID-19 epidemic spread in mainland China and its influencing factors. *Sci. Total Environ.* 744:140929. doi: 10.1016/j.scitotenv.2020.140929.

Youlden, D.R., Cramb, S.M., and BaadeP.D. (2008). The International Epidemiology of Lung Cancer: geographical distribution and secular trends. *J. Thorac. Oncol.* 3(8):819-831. doi: 10.1097/JTO.0b013e31818020eb.

Zheng, Z., Peng, F., Xu, B., Zhao, J., Liu, H., Peng, J., et al. (2020). Risk factors of critical & mortal COVID-19 cases: A systematic literature review and meta-analysis. *J. Infect.* 81(2):e16-e25. doi: 10.1016/j.jinf.2020.04.021.

Zou, H., and Hastie, T. (2005). Regularization and variable selection via the elastic net. *J. Roy. Stat. Soc. B* 67(2):301-320. doi: 10.1111/j.1467-9868.2005.00503.x.

## Tables

| Variable | Acronym | Transformation |
|---|---|---|
| Built-up area per capita | BUAPC | $(x - \min(x))^{1/3}$ |
| Urban population | UP | $x^2$ |
| Infant mortality | IM | $\log(x)$ |
| Gross domestic product per capita | GDPpc | $\log(x)$ |
| Human development index | HDI | $(\max(x) - x)^{1/2}$ |
| Number of immigrants minus emigrants | I-E | $(\max(x) - x)^{1/2}$ |
| Percentage of refugees | RE | $\log(x)$ |
| Average blood cholesterol level | CH | $(\max(x) - x)^{1/2}$ |
| Prevalence of obesity | OB | $(\max(x) - x)^{1/2}$ |
| Prevalence and severity of chronic diseases | CD | $x^{1/3}$ |
| Prevalence of insufficient physical activity | IN | $\log(\max(x) - x)$ |
| BCG immunisation coverage | BCG | $(\max(x) - x)^{1/2}$ |
| Epidemic onset | ON | $\log(x)$ |
| Long-term PM2.5 pollution | PL | $\log(x)$ |
| Precipitation | PR | $\log(x)$ |
| Wind speed | WS | $x^{1/3}$ |
| Basic reproduction number of SARS-CoV-2 | $R_0$ | $\log(x)$ |

**Table 1** Data transformations.

| Principal component | Coefficient | Standard error | P value |
|---|---|---|---|
| Demo PC1 | 0.15 | 0.03 | $10^{-6}$ |
| Demo PC2 | 0.02 | 0.03 | 0.4 |
| Demo PC3 | -0.00 | 0.03 | 0.97 |
| Demo PC4 | 0.06 | 0.03 | 0.03 |
| Demo PC5 | -0.05 | 0.03 | 0.08 |
| Demo PC6 | 0.02 | 0.03 | 0.6 |
| Demo PC7 | -0.12 | 0.03 | $10^{-4}$ |
| Demo PC8 | -0.00 | 0.03 | 0.9 |
| Demo PC9 | -0.10 | 0.03 | $4 \cdot 10^{-4}$ |

Root Mean Squared Error = 0.31
Adjusted R-Squared = 0.33
F-statistics vs. constant model = 7
P-value ~ $2 \cdot 10^{-8}$

**Table 2** Multiple linear regression for demographic principal components.

| Principal component | Coefficient | Standard error | P value |
|---|---|---|---|
| Meteo PC1 | -0.14 | 0.03 | $4 \cdot 10^{-5}$ |
| Meteo PC2 | 0.00 | 0.03 | 0.99 |
| Meteo PC3 | -0.04 | 0.03 | 0.3 |
| Root Mean Squared Error = 0.35<br>Adjusted R-Squared = 0.12<br>F-statistics vs. constant model = 6<br>P-value ~ $4 \cdot 10^{-4}$ | | | |

**Table 3** Multiple linear regression for meteorological principal components.

| Principal Component | Coefficient | Standard error | P Value |
|---|---|---|---|
| Demo PC1 | 0.13 | 0.03 | $2 \cdot 10^{-4}$ |
| Demo PC4 | 0.06 | 0.03 | 0.03 |
| Demo PC7 | -0.11 | 0.03 | $2 \cdot 10^{-4}$ |
| Demo PC9 | -0.10 | 0.03 | $9 \cdot 10^{-4}$ |
| Meteo PC1 | -0.04 | 0.03 | 0.3 |
| Root Mean Squared Error = 0.31<br>Adjusted R-Squared = 0.34<br>F-statistics vs. constant model = 13<br>P-value ~ $10^{-9}$ | | | |

**Table 4** Multiple linear regression for relevant principal components.